\documentclass[letterpaper]{article}

\usepackage[english]{babel}
\usepackage[margin=1in]{geometry} % Standard 1-inch margins
\usepackage{setspace}
\usepackage{helvet} 

\usepackage{graphicx}
\usepackage{caption} % adjust figure captions
\usepackage{array}
\usepackage{tabularx}
\usepackage{amsmath}
\usepackage{amssymb}
\usepackage{gensymb}

\usepackage[backend=bibtex, style=nature, sorting=none]{biblatex}
\bibliography{references}

\usepackage{xcolor} % Enable colored text
\usepackage{hyperref}
\hypersetup{
  colorlinks   = true,   % Use colored links
  urlcolor      = blue,   % URL links in blue
  linkcolor    = black,  % Internal links in black
  citecolor    = black   % Citation links in black
}

\usepackage{float} % Manage figure/table placement
\usepackage[all]{hypcap} % Anchor hyperlinks to top of figures

\usepackage{authblk}

\usepackage{verbatim}

\usepackage{adjustbox}

\renewenvironment{abstract}{%
        \begin{center}
        \begin{minipage}{14cm}
        {\textbf{\abstractname:}}
}{
        \end{minipage}
        \end{center}
}

\newcommand{\keywordsname}{Keywords}
\newenvironment{keywords}{
        \def\abstractname{\emph{\keywordsname}}
        \begin{abstract}}{\end{abstract}}

\title {Foundations of a Compositional Systems Biology}
\author{Eran Agmon \href{https://orcid.org/0000-0003-1279-2474}{\includegraphics[scale=0.04]{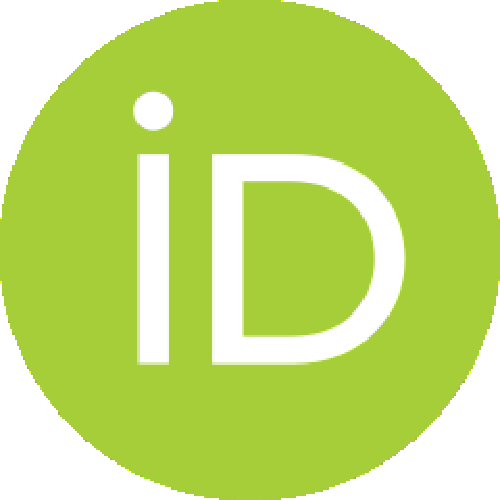}} \thanks{Correspondence address: agmon@uchc.edu } $^1$ }
\affil{$^1$Center for Cell Analysis and Modeling, Department of Molecular Biology and Biophysics, University of Connecticut Health, Farmington, Connecticut, USA.}
\date{}

\begin{document}

\maketitle
\vspace{4pt}

\begin{abstract}
Composition is a powerful principle for systems biology, focused on the interfaces, interconnections, and orchestration of distributed processes to enable integrative multiscale simulations. 
Whereas traditional models focus on the structure or dynamics of specific subsystems in controlled conditions, compositional systems biology aims to connect these models, asking critical questions about  \textit{the space between models}: 
What variables should a submodel expose through its interface?
How do coupled models connect and translate across scales?
How do domain-specific models connect across biological and physical disciplines to drive the synthesis of new knowledge?
This approach requires robust software to integrate diverse datasets and submodels, providing researchers with tools to flexibly recombine, iteratively refine, and collaboratively expand their models.
This article offers a comprehensive framework to support this vision, including: a conceptual and graphical framework to define interfaces and composition patterns; standardized schemas that facilitate modular data and model assembly; biological templates that integrate detailed submodels that connect molecular processes to the emergence of the cellular interface; and user-friendly software interfaces that empower research communities to construct and improve multiscale models of cellular systems. 
By addressing these needs, compositional systems biology will foster a unified and scalable approach to understanding complex cellular systems.
\end{abstract}


\vspace{4pt}

\section*{Introduction}
Cellular systems are multimodal and multiscale, with diverse mechanisms connecting across many levels of organization. 
Their collective behaviors, and connections to their environments, drive the evolution of dynamic self-organizing hierarchies—the biosphere being made of a bewildering number of cells that adapt, grow, and evolve within their environments, which drives the emergence of structures such as communities, tissues, organisms, ecosystems—and each cell itself a multilayered structure made of molecules, complexes, condensates, organelles. 
Modeling such systems demands a shift from examining individual subsystems in isolation to an integrative approach that emphasizes the interfaces of different subsystems and how they connect. 
%This integrative approach does not have to restrict to software solutions but also to the conceptual shift that helps us view biological systems as dynamic hierarchies, where different levels interact in meaningful ways.
The complexity of cells, together with the large number of researchers needed to thoroughly analyze even a single cellular subsystem, underscores the need for an open, collaborative scientific framework in which new data and models can be combined and recombined in the synthesis of new biological knowledge. 

Compositionality is here offered as an overarching principle to facilitate the integration of diverse datasets and models into open-ended simulations of cellular systems. 
This use of the term ``compositionality'' is adapted from category theory \cite{fong2019invitation} and software design \cite{gamma1995design}, but in this article reinvisioned for systems biology. 
Three fundamental criteria underpin compositionality. 
First, the \textbf{interfaces of subsystems}, which act as points of interaction. 
Second, a \textbf{composition pattern} linking different subsystems through their interfaces. 
Third, an \textbf{orchestration pattern} that drives composition in time by coordinating the subsystems' activities. 
For a complete definition of a composite system, each of these criteria requires a clear, standardized specification that enables model reproduction, sharing, iteration, and extension. 
A consistent set of interoperable process interfaces, composition patterns, and orchestration patterns will here be referred to as a \textbf{composition protocol}.

Composition pervades biology. 
At the molecular scale, cells are composed of a vast number of molecules that interact through their specific interfaces to drive changes to their structures, functions, and surroundings. 
The particular molecular composition of a cell determines its interface with the environment, through which it exchanges the material, energy, and information required to sustain itself and drive its growth. 
These interactions are constantly changing, as cells respond to signals, compete for resources, cooperate with other cells, and restructure their very environment. 
At higher levels of organization, populations of cells compose into communities, organisms, up to ecosystems. 
The patterns of how cellular interfaces connect with each other and with their environments determines their chemical, electrical, and mechanical inputs, driving their behavior, and setting the course of their evolution.

Systems biology models have not been compositional in the same way as their biological counterparts.
This is in large part because they do not have chemistry and physics to mediate their interactions, and require a model for every natural process.
It is true that some modeling paradigms such as ordinary differential equations (ODEs), flux balance analysis (FBA), Bayesian networks, rule-based models, and Markov models are individually compositional in that they can be expanded by adding new expressions or nodes. 
But different aspects of biology are typically studied with different modeling paradigms, and these too need to be brought together. 
The need to connect models across scales has driven interest in hybrid approaches that combine modeling paradigms—such as stochastic with deterministic \cite{haseltine2005origins}, kinetic ODEs with steady state FBA \cite{mahadevan2002dynamic,covert2008integrating}, particle-based with continuous spatial \cite{schaff2016numerical}, and whole-cell models with many interacting processes \cite{karr2012whole,macklin2020simultaneous} including kinetic models that connect to 3D spatial models with molecular resolution \cite{maritan2022building,thornburg2022fundamental,stevens2023molecular}. Foundational efforts like the Physiome Project have emphasized the importance of linking scales systematically to model the dynamics of biological systems from proteins to organs \cite{hunter2003integration}.
Multi-cellular models often use agent-based models (ABMs) \cite{walpole2013multiscale,ghaffarizadeh2018physicell,swat2012multi}—a class of hybrid models that span two scales, with a model of an environment and a model of individual agents. 
However, current hybrid models are usually built hard-coded in their own closed software ecosystem, without an easy way to flexibly expand and connect with external models that use different methods. 

\begin{figure}[H]
    \centering
    \includegraphics[width=0.9\textwidth]{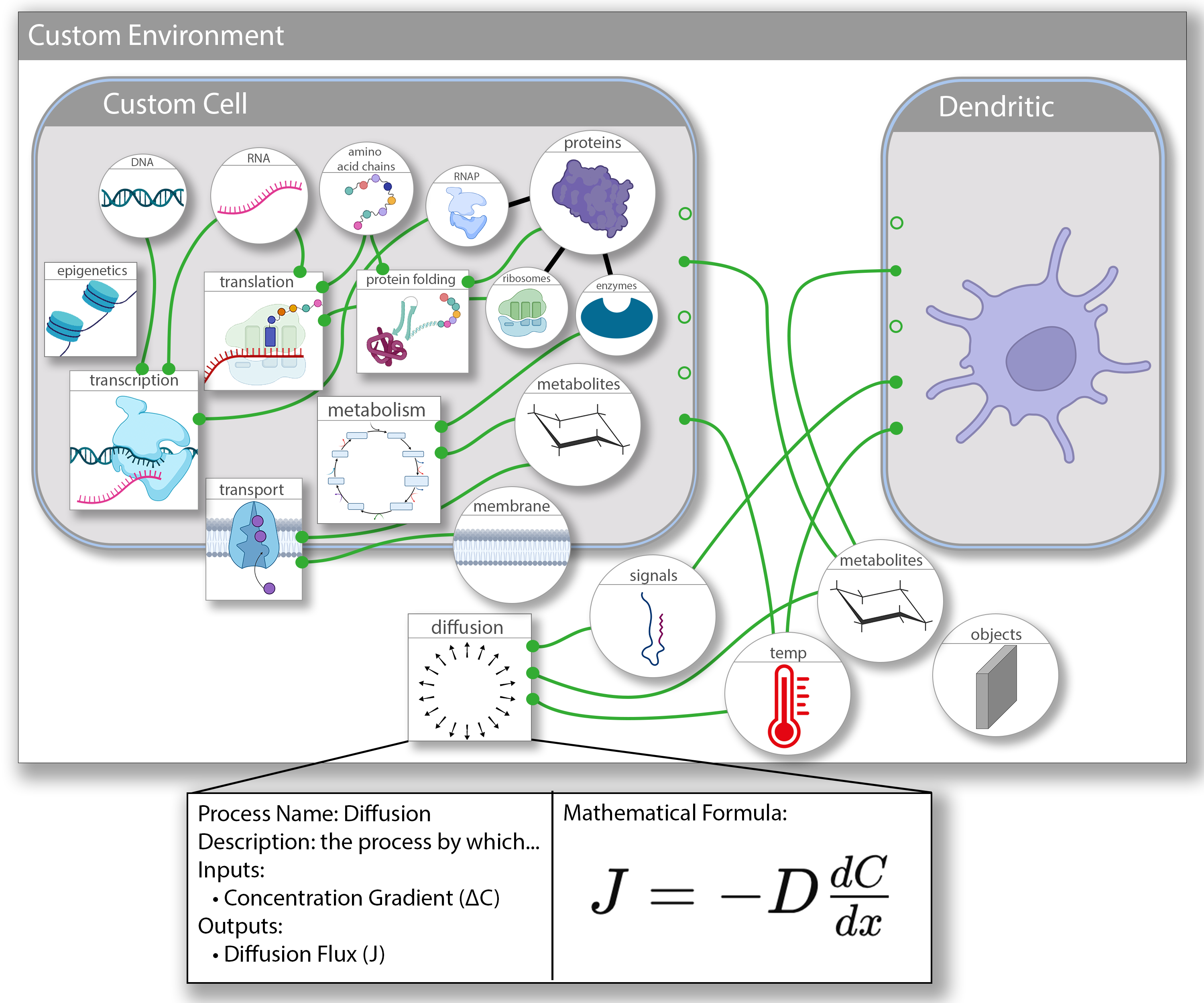}
    \caption{
    \textbf{Conceptual design of a user interface for compositional systems biology.}
    This illustrates how cellular systems can be modeled through the composition of submodels connected by wires. 
    The graphics display nested modules, with multiple cells within an environment, and within the custom cell are subcellular processes like metabolism, transcription, translation and protein folding. 
    The zoomed-in diffusion process displays model information. 
    Additional interactive elements may include clickable user elements to run the simulation, menus, and drag-and-drop interaction for intuitive model construction.
    Making this vision a reality requires a robust composition framework.
    }
    \label{fig:user_demo}
\end{figure}

To address the challenge of seamlessly integrating diverse models and methods into cohesive, multiscale simulations, we developed the Vivarium software \cite{agmon2022vivarium}. 
Vivarium has been applied to a range of domains, including the integration of FBA with kinetic ODEs and agent-based modeling \cite{agmon2022vivarium}, bacterial chemotaxis that combines motile forces with a chemical reaction network \cite{agmon2020multi}, agent-based modeling of tumor morphogenesis integrated with multiplexed spatial imaging \cite{hickey2024integrating}, physical models of actin at multiple scales \cite{vegesna2024comparing}, and a comprehensive whole-cell model of \textit{E. coli} that combines FBA, ODEs, and stochastic algorithms \cite{skalnik2023whole}.
Adapting earlier work on modular simulation frameworks \cite{ginkel2003modular,eker2003taming,chopard2014framework,veen2020easing,blochwitz2012functional}, Vivarium revitalizes the application of such efforts to systems biology by introducing a hierarchical structure and process-centric design that not only unifies multiscale \cite{dada2011multi} and middle-out \cite{walker2009virtual} approaches but also facilitate the integration of previously isolated simulation tools \cite{shaikh2022biosimulators}.
Figure \ref{fig:user_demo} illustrates a conceptual design of a possible user interface for Vivarium, showcasing how cellular systems can be modeled through the composition of submodels connected by wires, and how this structure provide intuitive integration across scales.

Vivarium's composition framework, called process bigraphs, is being extended with the process-bigraph software suite—to be introduced in a separate more technical paper so that this current article can focus on composition in a more general manner.
This article outlines the core requirements of a compositional systems biology and proposes its initial instances with templates for molecular and cellular compositions. 
This provides a foundation for robust simulation infrastructure, and also useful conceptual tools for thinking about cellular systems.
These topics are described in the following sections:
\begin{itemize}
    \item \textbf{\hyperref[sec:composition_framework]{Composition framework}.} Introduces the basic concepts of process bigraphs, how interfaces work, how they connect in nested composite systems, and how they can be orchestrated to add/remove, connect, and simulate processes at different scales. 
    Rather than going into the framework’s formal underpinnings, diagrams are used to graphically represent the concepts—these  provide a visual language that will be reused to build more complex concepts throughout the manuscript. 
    This section also recommends a standardized protocol for shared infrastructure that can support an open-ended community-driven modeling.
    \item \textbf{\hyperref[sec:templates_for_biology]{Templates for multiscale cellular modeling}.} Looks to biology to determine the cellular interface, and analyze how this interface can emerge from the composition of molecular mechanisms. 
    This provides a compositional perspective on the emergence of cells from self-organizing molecular processes, how they navigate viability, how they grow, divide, develop into multilayered dynamic organizations, how they ultimately disintegrate back into molecules, and how their ongoing orchestration drives evolution. 
    Biological schemas specified as process bigraphs can be thought of as templates for a longer-term goal, to fill in detailed mechanisms that through co-simulation will recapitulate the behavior of real biological cells. 
    \item \textbf{\hyperref[sec:collaborative_biosciences]{Collaborative biosciences}.} The discussion focuses on the impact compositional thinking can have on systems biology as a field. 
    This focuses on knowledge integration and the feedback between models and their corresponding real-world biological systems, as mediated by researchers through intuitive software interfaces and collaborative research practices. 
\end{itemize}
By emphasizing the composition of data, models, schemas, software, and research efforts, we are taking important steps towards a more compositional systems biology. 
Such a framework not only facilitates the integration of diverse biological data and models but also ensures that these drive the synthesis of new biological knowledge.

%%%%%%%%%%%%%%%
% Composition Framework  %
%%%%%%%%%%%%%%%

\section*{Composition framework}
\label{sec:composition_framework}

To ensure the ongoing scalability of systems biology models, we need to design a robust and general-purpose framework that can scale to any biological problem. 
This would need to be easy to use and compatible with any modeling effort across systems biology. 
Key features of a composition framework include standardized \textbf{process interfaces}, \textbf{composition patterns}, and \textbf{orchestration patterns}, which give researchers the ability to define a biological function, determine how it connects with external states and how it is generated by internal processes. 
This section offers a high-level overview of these requirements, setting aside a detailed description of the formal system for another paper, and instead relying on \textbf{composition diagrams} to visually represent the different components of a compositional model, including processes, ports, states, connections, and nesting. 

\begin{figure}[H]
    \centering
    \includegraphics[width=1.0\textwidth]{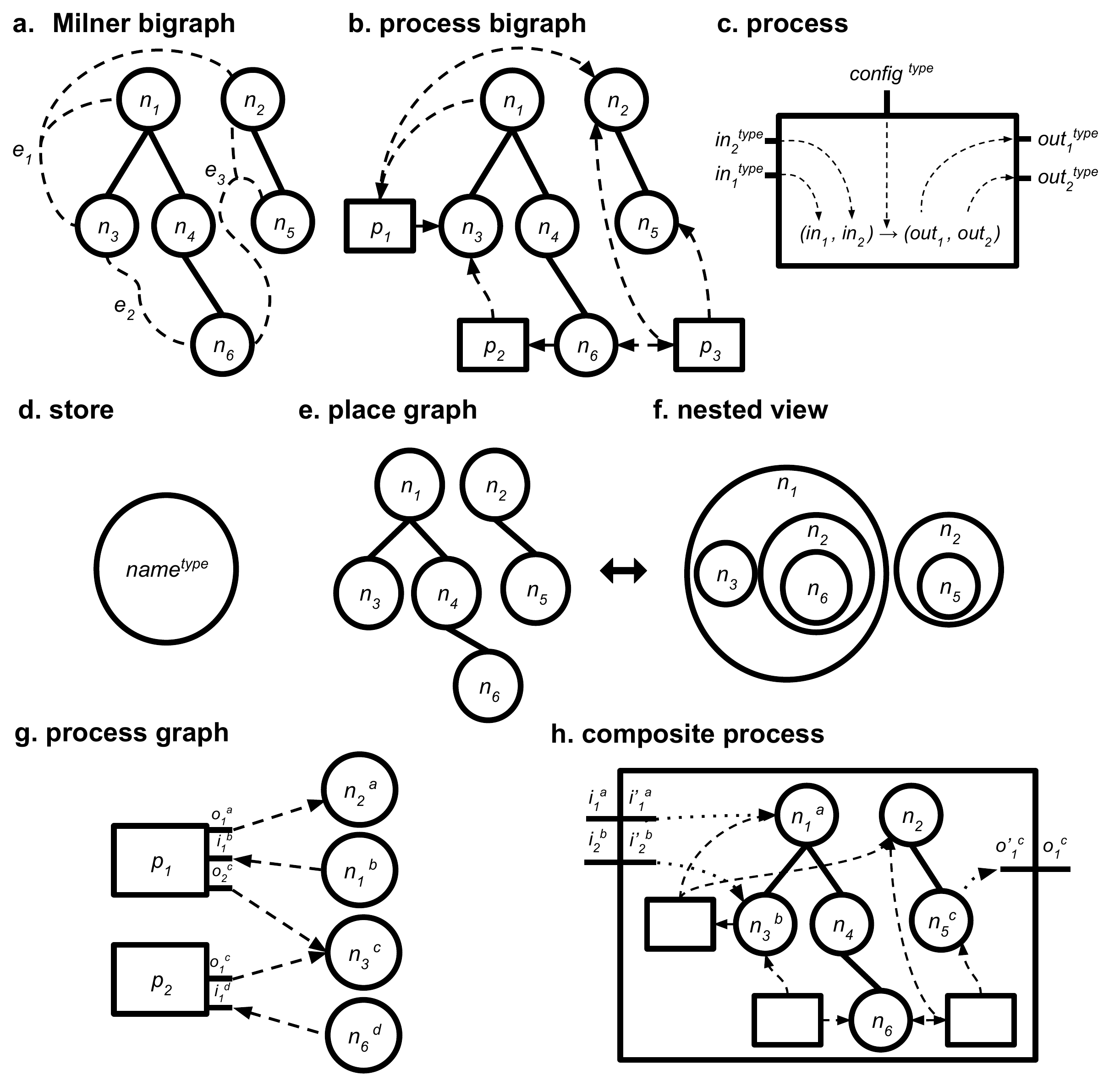}
    \caption{
    \textbf{Composition framework overview.}
    \textbf{a.} The original "Milner" bigraphs consist of a link graph with hyperedges ($e_1, e_2, e_3$) and nodes ($n_1, n_2, n_3, n_4, n_5, n_6$), and a place graph connecting the nodes (see \ref{fig:composition_framework}f).
    \textbf{b.}  Process bigraphs replaces the link graph with a process graph, made of processes  ($p_1, p_2, p_3$) connecting to the nodes via their ports. 
    \textbf{c} A process is depicted as a rectangle with ports along its boundary to represent its interface. 
    Input ports are on the left and outputs on the right (though this arrangement is not enforced).
    Ports specify the type of information flow in and out, shown as superscript (a generic ``type'' in this figure). 
    The process's update function is a mapping from inputs and outputs, informed by its config. 
    \textbf{d.} Stores, depicted as circles, hold any data type and get nested in hierarchies for multiscale representation. 
    Each store shows its name (e.g., ``name'') and the data type it holds (here, ``type'').
    \textbf{e.} A place graph of nested stores shows outer stores connected to inner stores by thick black edges.
    \textbf{f.} Nested view of the place graph, with inner stores shown within their outer stores.   
    \textbf{g.} A process graph consists of processes connected to stores via ports. 
    Port types must match the connected store type (types \textit{a, b, c} shown here), with inputs/outputs indicated by arrow directions. 
    \textbf{h.} A composite process contains an internal process bigraph that it orchestrates. 
    The composite has external input/output ports that can connect to external stores, with matching internal ports linked to the internal bigraph via dotted wires.
    Variables connected across these inner/outer port pairs can pass updates to remain synchronized.
    }
    \label{fig:composition_framework}
\end{figure}

This article focuses on a composition framework called ``process bigraphs'', which is an evolutionary extension of Robin Milner’s \textbf{bigraphs} \cite{milner2009space} (Fig. \ref{fig:composition_framework}a), and was initially implemented by Vivarium \cite{agmon2022vivarium}. 
Bigraphs are a powerful framework for compositional modeling due to their ability to represent complex systems through hierarchical structures and flexible reconfigurations, related to rule-based and agent-based modeling. 
A bigraph combines two distinct graph structures: a \textbf{place graph} (shown in Fig. \ref{fig:composition_framework}e), which represents the hierarchical nesting of entities within other entities, and a \textbf{link graph}, which represents the connectivity of a system using hyperedges to capture the communication between multiple entities. 
\textbf{Process bigraphs} (Fig. \ref{fig:composition_framework}b) reimagine this structure with the introduction of processes (Fig. \ref{fig:composition_framework}c), replacing the link graph with a process-state interaction graph, more simply called a \textbf{process graph} (Fig. \ref{fig:composition_framework}g). 
Processes introduce additional considerations for orchestration—the ordering of distributed events in time—synthesizing concepts from discrete-event co-simulation, functional programming, and distributed concurrent systems \cite{lamport2019time,thompson2009unifying,goldstein2018multiscale}.

% Process interface
\subsection*{Process interface}
Processes are the operational units of integrative simulations, each simulating distinct mechanisms that drive changes to the system. 
By emphasizing the \textbf{process interface} with a process-centered design, we enable modelers to scrutinize individual processes independently prior to their systematic integration in a broader ecosystem and prior to determining their particular implementation details. 

A process interface defines the types of state variables that are exposed by the process—its inputs, outputs, and configuration settings. 
This interface serves as a process type, enabling processes conforming to the same interface to be substituted within a composite system—similar to how enzymes catalyzing the same reaction can be interchanged. 
Interfaces may represent specific biological phenomena, such as a gene regulation process that maps transcription factors to gene expression outputs. 
Processes can operate at varying scales, from atomistic to molecular, and can be swapped based on the required level of detail.

Fig. \ref{fig:composition_framework}Cc depicts a process as a rectangle with input and output ports along its boundary, representing its interface.
In Equation \ref{eqn:process_interface} a process is defined by an update function that converts input data into output data. 
Each process is configured by a ``config'', a structured data element defined by a schema that includes constants, datasets, or model files that specify parameters or constraints for the process's operation.
The update function takes inputs through the ports, applies transformations based on the config, and outputs the resulting changes to states via the output ports. 
Both ports and configs are associated with types that determine the permissible values, ensuring compatibility with connected processes or stores. 
Ideally, the update function is a pure function—one that depends solely on its inputs and config and does not maintain internal state—allowing modular integration with other processes.

The process interface is defined by its components: the config, input and output ports (with their associated types), and the update function. This interface can be formally represented as:
\begin{equation}
\text{process}(\text{config}^\text{type}): (\text{in}_1^\text{type}, 
\text{in}_2^\text{type}) \rightarrow (\text{out}_1^\text{type}, 
\text{out}_2^\text{type})
\label{eqn:process_interface}
\end{equation}

By providing a general interface with standard methods of composition, we open the framework to many possible methods—it could support a simulation module, configured by passing in parameters or model file, receiving initial conditions through the input ports and returning simulation results as an output. 
It could be a translator, which reads an input dataset and converts it to a format understood by a different simulator. 
It could be a cell agent, with a specific cellular phenotype specified in the config, and an update function that maps its sensory inputs to behavioral outputs. 
It could be a neural network, configured by weights and meta-parameters. 
Or a figure generator that reads the simulation state as input, and returns a rendered figure as output.
The options are endless, keeping the framework open-ended to evolve with user needs.

\textbf{Stores} (Fig. \ref{fig:composition_framework}d) facilitate interactions between processes by holding and managing the externalized variables, and serve as conduits through which processes connect. 
Graphically represented as circles, stores are nodes that hold values defined by data types. 
In a type system, each type dictates how processes can interconnect; processes can only link to a store through a port that share the same type, ensuring coherence. 
Types may capture information such as units (Joules, moles, meters), data types (integers, arrays, dictionaries), or more complex types like position coordinates (x, y, z) or images.
By reporting the type of our datasets, we can implement accurate connections and facilitate automatic composition. 
Stores can also be nested within each other using a place graph (Fig. \ref{fig:composition_framework}e,f)—sometimes called a tree or forest graph—this format facilitates the structuring complex biological data into a format that reflects the hierarchical organization of biological states.

% Composition patterns
\subsection*{Composition patterns}
Composition patterns, or ``wiring'', details how process connect to each other through their ports to shared stores, establishing relationships between their interfaces. 
When connecting stores and processes at the same level, without nesting, wiring determines a pure \textbf{process graph} in which multiple processes can connect to stores through their ports with matching types (Fig. \ref{fig:composition_framework}g). 
When there is a hierarchy, processes can connect across levels, translating states at the micro-level to the macro-level, or setting top-down constraints—the composition of a place graph and process graph is the process bigraph (Fig. \ref{fig:composition_framework}b).
%Distinct processes that declare the same interface can be interchanged to fulfill these relationships, thus allowing the simulation engine to swap out processes with comparable interfaces. 

A composite, itself made of multiple processes and stores, can itself be used as process (Fig. \ref{fig:composition_framework}h).
This means a composite has its own interface and can itself function within a higher-order composite.
To achieve this requires inward-facing ports that connect to the internal stores—the dotted-line wires in Fig. \ref{fig:composition_framework}h—which synchronize to their corresponding external store across the interface. 
A composite needs to implement an orchestration pattern that determines the order by which it triggers internal processes (discussed in \ref{sec:orchestration}). 
The possibility of composite processes means we can run full hybrid simulations and plug them together as modules of a super-simulation — an integrative multiscale simulation \textit{within} an integrative multiscale simulation that can run on a separate computer and only synchronize the required states across the composite interface. 
Because of this property, process-bigraphs can replace simple processes with composite processes that have matching interfaces—equivalent to zooming into composite processes to reveal their internal structure and function.
This could also be the basis of a coarse-graining strategy.

% Orchestration patterns
\subsection*{Orchestration patterns}
\label{sec:orchestration}
Composite simulations host numerous interacting processes that operate concurrently and across timescales, each fulfilling unique functions. 
These processes are managed by the composite’s update function—its simulation engine—through \textbf{orchestration patterns}, which determine their ordering, triggering, and result-gathering. 
Composites act as orchestrators, following their orchestration patterns to selecting which processes to run, projecting the simulation states to their ports, triggering their updates, collecting the updates and applying them to the internal state.
This dynamic orchestration creates a flexible framework where processes can be added, removed, rewired, or even reconfigured during runtime, enabling the emergence of new structures and behaviors.

\begin{figure}[H]
    \centering
    \includegraphics[width=1.0\textwidth]{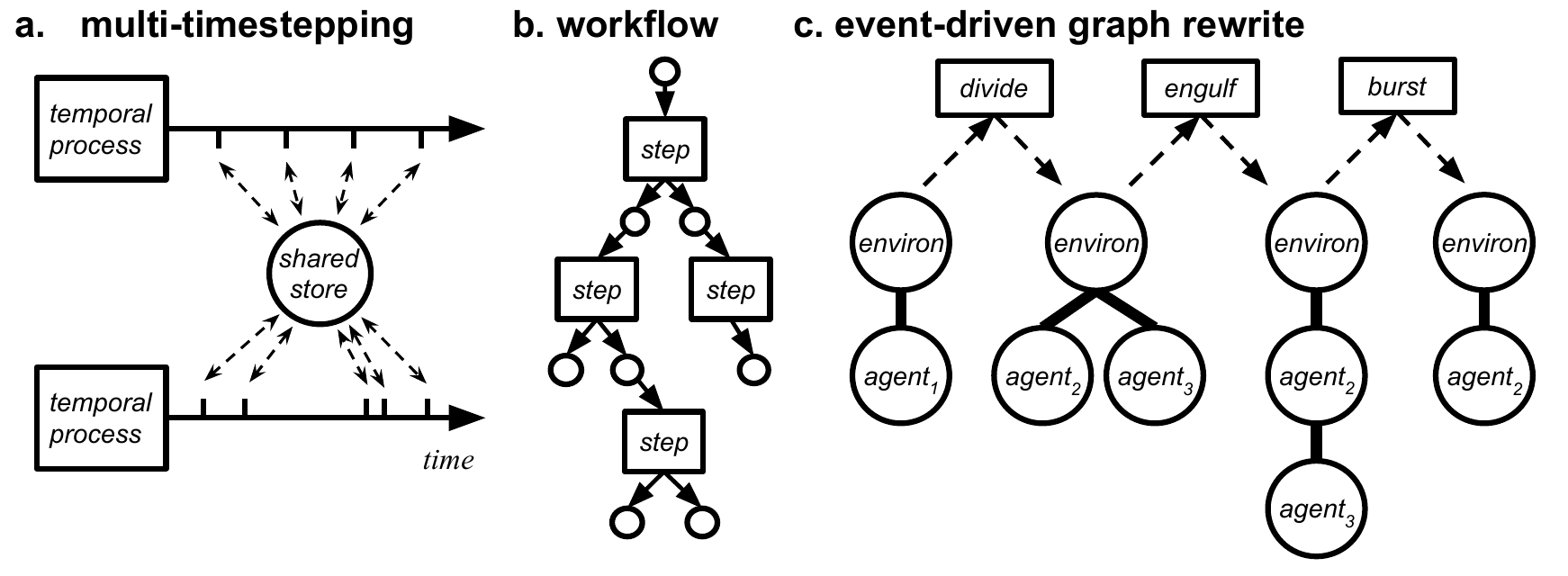}
    \caption{
    \textbf{Orchestration patterns.}
    \textbf{a.} Multi-timestepping, with temporal processes each updating at their preferred time intervals, orchestrated by a discrete-event co-simulation engine. 
    \textbf{b.} A workflow is a directed acyclic graph that sets the order of updates for step processes, with each one triggered by changes to its input states. 
    \textbf{c.} Event-driven graph re-writes orchestrated as instantaneous events, change the topological structure of an agent-environment system. 
    Each graph re-write can be specified as a reaction, triggered by states internal or external to the agents in the environment. 
    ``Divide'' makes one agent divide into two. 
    ``Engulf'' places one agent within the other. 
    ``Burst'' dissolves the agent, releasing its inner components back to the environment.
    }
    \label{fig:orchestration_diagram}
\end{figure}

There are several useful orchestration patterns included in Vivarium, illustrated in Figure \ref{fig:orchestration_diagram}. 
\textbf{Multi-timestepping} (Fig. \ref{fig:orchestration_diagram}a): asynchronous processes update at their preferred timescales. 
They are managed by a discrete-event co-simulation method that schedules the time-to-next update, collects the results, and triggers each process to run in parallel.
\textbf{Workflows} (Fig. \ref{fig:orchestration_diagram}b) use directed acyclic graphs (DAGs) to set the order of process updates, with each triggered by the completion of preceding processes, or by changes to upstream stores. 
\textbf{Event-driven graph rewrites} (Fig. \ref{fig:orchestration_diagram}c), sometimes called \textit{reactions} or \textit{rules}, are updates that restructure the composite’s graph—this can include moving, adding, and removing stores or processes, or rewiring process process ports to different stores.
Finally, different orchestration patterns can be composed and made interoperable, for example having workflows and graph rewrites running between temporal processes, or entire integrative simulations running as a step of a workflow. 

Other methods of orchestration can include the management of ensemble models, with many parallel simulation of model variants being launched in parallel, and their results connected into a unified prediction.
Biomedical digital twins \cite{laubenbacher2022building} require ongoing streaming input, with orchestration that can continually update the composite's parameters, processes, and overall structure to match observations as they become available.
Discovery engine composites may run inference on their internal models, and through their output ports report new parameter values or other inferred knowledge.

% Composition protocol
\subsection*{Standardized composition protocol}
\label{sec:standard_composition_protocol}

A standardized composition protocol, built on shared schemas for process interfaces, composition patterns, and orchestration patterns, can establish the foundation for  systems biology. 
Inspired by the role of TCP/IP in the internet, this protocol would aim to standardize communication between process models, software modules, databases, and other applications including web interfaces, AI assistants, and experimental protocols for high-throughput data streaming. 
Just as TCP/IP created a dynamic, expandable network open to anyone's contributions, this framework could build a scalable and interoperable ecosystem of simulation tools for building community-driven compositional models.

Schemas within this framework provide structured data formats that ensure reproducibility, compatibility across tools, and ``plug-and-play'' integration of new processes and datasets into existing simulations. 
By aligning with existing standards such as SBML or CellML for biological networks \cite{keating2020sbml,lloyd2004cellml}, spatial model formats \cite{schaff2023sbml}, and multi-cellular models \cite{fletcher2022seven}, alongside resources like the BioModels database \cite{le2006biomodels}, BioSimulators database \cite{shaikh2022biosimulators}, and SimService for containerized modules \cite{sego2024simservice}, the protocol creates a unified framework for sharing and combining simulation modules. 
By leveraging these shared standards, tools ranging from databases to user interfaces can connect seamlessly, fostering collaboration across disciplines and teams.

This protocol advances the FAIR (Findable, Accessible, Interoperable, Reusable) principles \cite{wilkinson2016fair} by ensuring tools and datasets report their interfaces and type information, enabling researchers to reliably share and combine simulation modules. 
By setting explicit types on process ports, schemas define the kinds of information exchanged between modules, ensuring consistency and seamless communication. 
Structured schemas also support robust annotation, linking experimental data to simulation components and preserving data provenance throughout the modeling pipeline. 
This enhances transparency and reproducibility, allowing researchers to trace how data informs models and integrate diverse sources effectively. Table \ref{tab:composition} illustrates these principles with schemas for process interfaces and compositions, showing how inputs, outputs, and configurations of processes can connect through shared “wires” to build modular and scalable hybrid models.

By uniting tools and efforts through shared schemas, this approach transforms systems biology into a dynamic and collaborative ecosystem. 
Researchers can more effectively integrate diverse models and orchestrate simulations, driving innovation and enabling solutions to complex biological challenges that were previously out of reach.

\begin{table}[H]
\centering
\renewcommand{\arraystretch}{1.2} % Adjust row height
\setlength{\tabcolsep}{4pt} % Adjust column padding
\begin{adjustbox}{max width=\textwidth,center}
\begin{tabular}{|p{0.15\textwidth}|p{0.4\textwidth}|p{0.4\textwidth}|}
\hline
 & \textbf{Schema} & \textbf{Diagram} \\ \hline

\textbf{Interface} & 
\begin{minipage}[t]{\linewidth}
%\footnotesize % Smaller font size for compactness
\texttt{metabolism:}\\
\hspace*{5mm}\texttt{type: flux balance}\\
\hspace*{5mm}\texttt{config: fba}\\
\hspace*{5mm}\texttt{inputs:}\\
\hspace*{10mm}\texttt{- nutrients: array[mol]}\\
\hspace*{10mm}\texttt{- enzymes: array[mol]}\\
\hspace*{5mm}\texttt{outputs:}\\
\hspace*{10mm}\texttt{- fluxes: array[mol/s]}\\
\end{minipage} & 
\adjustbox{valign=t, max height=5cm, max width=6cm}{\includegraphics{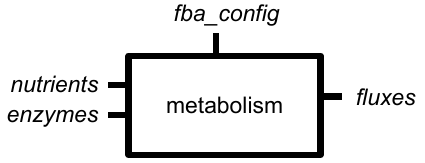}} \\ \hline

\textbf{Composition} & 
\begin{minipage}[t]{\linewidth}
%\footnotesize
\texttt{metabolism:}\\
\hspace*{5mm}\texttt{address: local:COBRA}\\
\hspace*{5mm}\texttt{config: ecoli\_metabolism}\\
\hspace*{5mm}\texttt{wires:}\\
\hspace*{10mm}\texttt{- fluxes: metabolites}\\
\hspace*{10mm}\texttt{- nutrients: metabolites}\\
\hspace*{10mm}\texttt{- enzymes: proteins}\\
\texttt{gene expression:}\\
\hspace*{5mm}\texttt{address: local:COPASI}\\
\hspace*{5mm}\texttt{config: ecoli\_genes}\\
\hspace*{5mm}\texttt{wires:}\\
\hspace*{10mm}\texttt{- proteins: proteins}\\
\hspace*{10mm}\texttt{- genes: genes}\\
\end{minipage} & 
\adjustbox{valign=t, max height=6cm, max width=6cm}{\includegraphics{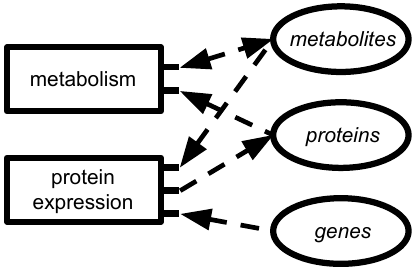}} \\ \hline

\end{tabular}
\end{adjustbox}
\caption{
\textbf{Demonstration of the composition protocol through example schemas and diagrams for a process interface and composition.} 
This table illustrates the basic capabilities of a standard, showcasing how process interfaces define configurations, inputs, and outputs with explicit type information (arrays of moles, arrays of molar flux). 
The composition schema demonstrates how these interfaces are wired to shared stores, enabling modularity and interoperability between submodels. 
Note that this is not the finalized standard but serves as a conceptual framework to highlight the potential of a standardized approach to compositional modeling.
}
\label{tab:composition}
\end{table}

%%%%%%%%%%%%%
% Biological Templates  %
%%%%%%%%%%%%%

\section*{Templates for multiscale cellular modeling}
\label{sec:templates_for_biology}

With the conceptual and graphical framework for composition in place, we can begin our compositional analysis of cells—a similar approach could be taken for different biological scales, such as tissues, multicellular organisms, or whole ecosystems. 
A compositional analysis of cells requires taking a dual perspective, of the cell as a process interfacing an environment, and of the cell as a composition of interacting molecules. 
By taking these perspectives simultaneously, we ask: 
How does the cellular interface emerge from interactions in the molecular domain? 
How is this interface maintained and remodeled by processes internal to the cell? 
How does this interface respond to environmental changes? 
What happens to this interface upon cell death? 
Growth and division introduce additional nuance, driving multi-cellular compositions to build up into heterogeneous multi-layered structures that develop and evolve—compelling us to ask how composition redefines the resulting dynamics.

While this section does not present quantitative models, it offers high-level templates for such models in the form of compositional diagrom of different cellular systems and the connections between them. 
Templates aim to guide the construction of more detailed models by first defining their input/output relations and translations, providing placeholders are only afterwards filled with detailed mechanisms and fit to experimental observations. 
By examining these templates through the lens of compositional principles, we gain insights into how systems interconnect, function, and evolve, revealing mechanisms that transform molecular interactions into organized, adaptable biological structures.

% Cellular interface
\subsection*{Cellular interface}

\begin{figure}[H]
    \centering
    \includegraphics[width=1.0\textwidth]{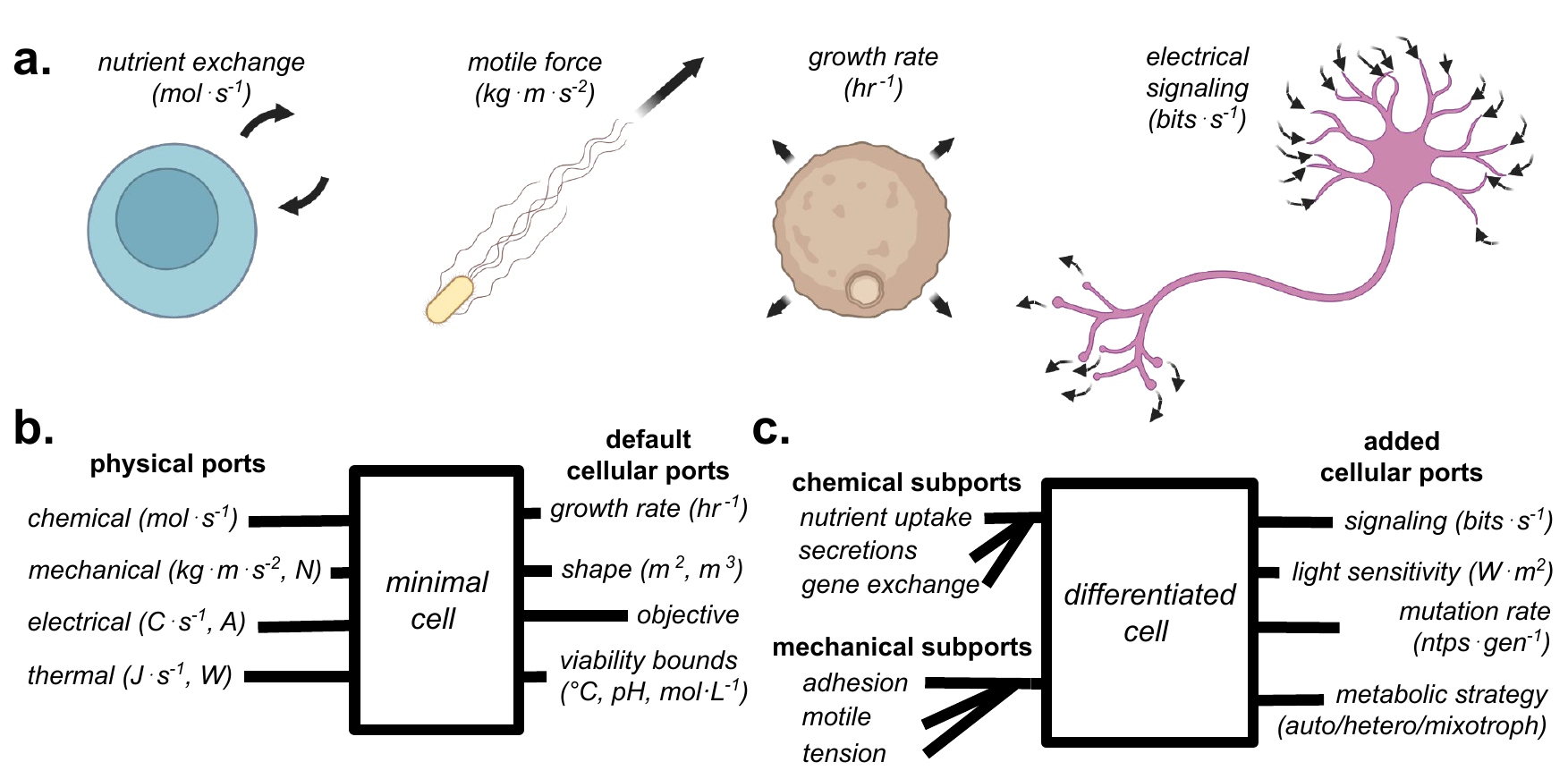}
    \caption{
    \textbf{Illustration of cellular interfaces and their ports.}
    \textbf{a.} Examples of physical and biological interactions represented as ports: chemical transport for molecular exchange, motile forces generated by structures like flagella, growth rate as a measure of cellular reproduction, and electrical signaling in excitable cells like neuron.
    \textbf{b.} The minimal cellular interface integrates physical interaction, including chemical, mechanical, electrical, and thermal exchanges, with cellular ports like such as growth rate, shape, objectives, and viability bounds.
    \textbf{c.} A more differentiated cellular interface refines existing ports and adds new ports. The chemical fluxes port is subdivided into nutrient uptake, secretions, and gene exchange. The surface forces port includes adhesion, motility, and tension molecules. New ports include light sensitivity and signaling are added.
    }
    \label{fig:cell_interface}
\end{figure}

Distinguishing a cell from its environment is easy to do observationally—every cell has a membrane that separates it from its surroundings. 
The membrane serves as an interface through which the cell and environment interact. 
These interactions may include the transport of molecules through channels, the binding of signals to membrane-bound proteins, the maintenance of a voltage gradient across the membrane, and the exertion of cytoskeletal forces that stretch and shape the membrane in response to environmental forces. 
Characterizing this interface requires defining its inputs and outputs (Fig. \ref{fig:cell_interface}a) with explicit type information, such as units, to quantify the variables connecting the cell to its environment.

The cellular interface embodies modularity, linking internal processes to external dynamics through defined ports—chemical, mechanical, electrical, and thermal—that serve as universal types across diverse contexts.
Key physical variables include chemical fluxes across the membrane (\(\text{mol} \cdot \text{s}^{-1}\)), mechanical forces (\(\text{m} \cdot \text{s}^{-2}\)), electrical currents (\(\text{C} \cdot \text{s}^{-1}\)), thermal energy transfers (\(\text{J} \cdot \text{s}^{-1}\) or watt, \(W\)) to describe the flow of matter and energy.
These interactions adhere to internal constraints like mass balance, flux balance, and entropy production, ensuring compositional consistency with physical laws and potentially informed by bond-graphs \cite{gawthrop2021modular}.
These constraints reinforce compositional consistency, ensuring that internal and external dynamics are coherently integrated across the interface.

Beyond physical variables, biological processes such as shape maintenance, signaling, and multicellular communication enrich the interface. 
Specialized components—adhesion molecules, receptors, and pumps—extend the interface's functionality, supporting more complex behaviors. 
Previous efforts like the Cell Behavior Ontology \cite{sluka2014cell} and MultiCellDS \cite{friedman2016multicellds} have characterized these elements but often blur distinctions between internal cellular processes and multicellular interactions. 
Compositional systems biology can refine these descriptions into precise, composable type systems that support multicellular modeling and expand the functionality of the minimal interface.

We should begin efforts with a foundation model that focuses on defining a core set of interface elements that are ubiquitous across cell types. 
We can consider the most basic cellular interface—a ``minimal cell interface'' that can serve as the foundation for more differentiated models (Fig. \ref{fig:cell_interface}b). 
This minimal interface may represent a common ancestral form of cells, which included the set of necessary processes that allowed early cells to survive. 
We could also focus on a known model microbe, such as \textit{Mycoplasma} or \textit{E. coli}, which can be the focus of initial development and then extended to other cellular models.
The minimal cellular interface can be thought of as a schema that all cell models should inherit from, connecting with the environment through ports for chemical fluxes, cell mass, and cell shape. 

Differentiated interfaces (Fig. \ref{fig:cell_interface}c) build on this foundation by refining existing ports or adding new ones, such as light sensitivity, genetic exchange, electrical conductance, and signaling. 
Specialized molecules, including adhesion proteins (cadherins, integrins) and motility generators (myosins, flagella), instantiate these functions. 
Flux ports can be subdivided into specific categories, such as nutrient transporters (glucose transporters), waste removal mechanisms (exocytosis, efflux pumps), and signal transduction (G-protein-coupled receptors, receptor tyrosine kinases). 
 This differentiation reflects the principle of functional extension, where core compositional elements are expanded into specialized systems to meet evolving requirements.
By systematically defining these interface port types, we transform the cellular interface into a robust, composable model that lays the groundwork for future refinements and helps decipher the evolutionary expansion of cellular interfaces.

Cellular models can also be configured with goals and viability boundaries. 
Goals are set as objectives, such as maximizing growth, maximizing production of a particular signal, or maximizing a different reward function. 
By providing these goals in our models, they can then be optimized through different internal methods, such as linear optimization, gradient descent, or genetic algorithms, to make predictions about what an optimal cell will do under different conditions. 
Viability bounds—sometimes called ``essential variables'' \cite{ashby2013design}—define the range of environmental conditions within which the cell can survive. 
These can apply to variables such as temperature (20–40\degree C), pH (6.5-7.5), osmolarity (e.g., ranges of solute concentration, \(\text{mol} \cdot \text{L}^{-1}\), and nutrient availability (e.g., minimum concentrations of glucose or nitrogen sources). 
If these bounds are crossed, the cell dies, and as a result, the cellular interface ceases to be relevant.

% Connecting to an environment
\subsection*{Connecting to an environment}
\label{sec:connecting_environment}

To accurately model a cell, it is essential to include the environmental processes that define its context, whether in controlled experimental setups, natural ecosystems, or physiological conditions. 
Connecting to an environmental state imposes specific potentials on the cellular interface, shaping the scope of information the cell can sense and respond to. 
The connection with an environment determines potentials across the cell interface, including chemical potential gradients (\(\text{J} \cdot \text{mol}^{-1}\)), electrical potentials (e.g., voltage across membranes in millivolts, mV), mechanical stresses (e.g., \(\text{N}  \cdot \text{m}^{-2}\)), and thermal gradients (e.g., temperature differences in \(\text{C}\degree\)). 
Together, they determine the information available for cellular processes to make adaptive decisions about their survival.

\begin{figure}[H]
    \centering
    \includegraphics[width=1.0\textwidth]{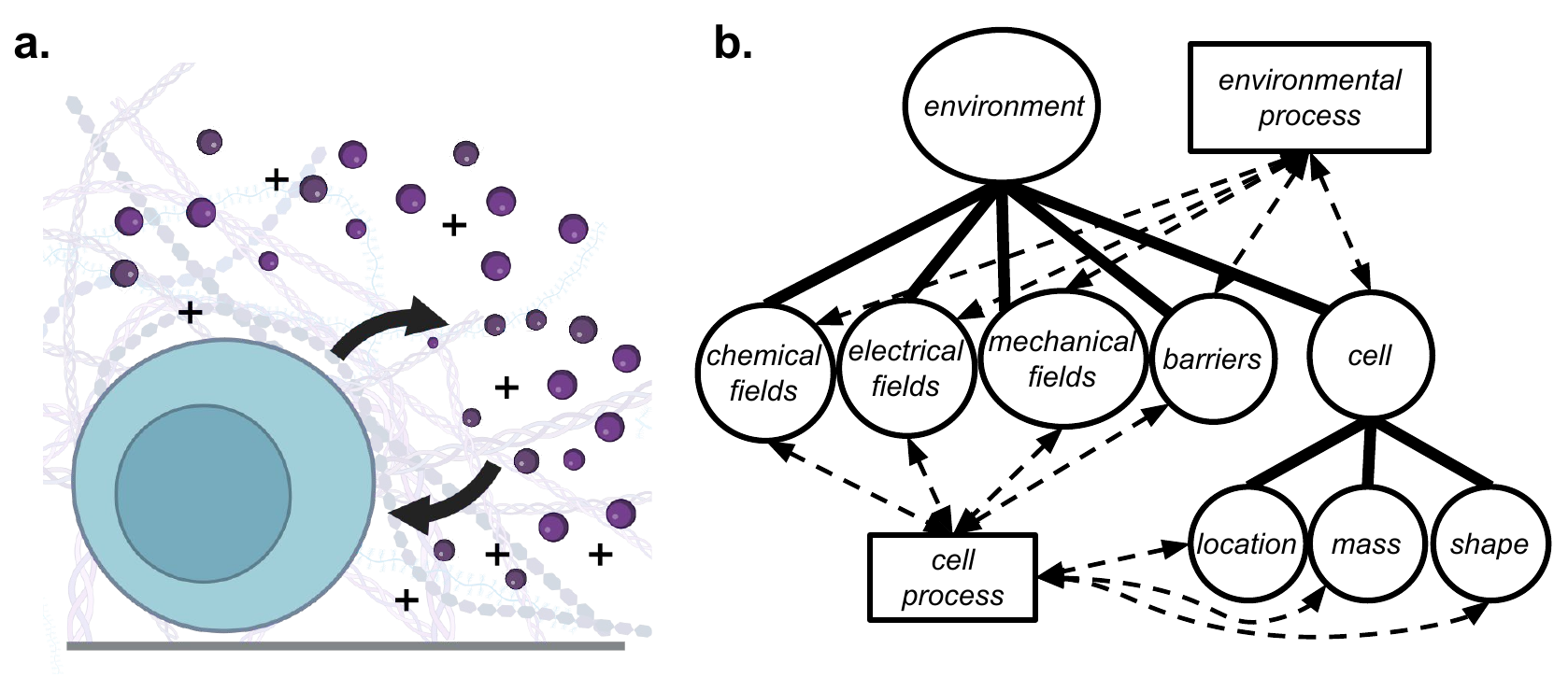}
    \caption{
    \textbf{Cell-environment composition.}
    \textbf{a.} Illustration of a minimal cell interacting with environmental states including molecules, electric charge, an extracellular matrix, and obstacles such as a floor. 
    %Environmental states may be impacted both by environment processes and cellular processes. 
    \textbf{b.} Composition diagram of a cell process mediating interactions between a cellular state (location, mass, shape) and an environmental state through exchanges with chemical, electrical, and mechanical fields, and with barriers. 
    An environmental process are shown connected to the environmental states, this represents many possible processes such as diffusion, advection, and other physics which may drive the environmental fields and barriers.
    }
    \label{fig:cell_environment}
\end{figure}

In simulations, environmental states are often represented as external fields of molecular concentrations or discrete objects that impose physical constraints, such as mechanical barriers or adhesive surfaces (Fig. \ref{fig:cell_environment}). 
Non-spatial models may simplify these states as resource pools. 
Environmental processes such as diffusion and advection govern the distribution and transport of molecules, shaping the availability of nutrients, signaling molecules, and waste removal near the cell. 
Mechanical processes can create stresses that deform the cell or influence its ability to adhere, migrate, or divide. 
Thermal processes, such as heat conduction or localized temperature changes, can affect reaction rates and metabolic efficiency. 

Just as the environment can influence a cell’s behavior, the cell can, in turn, modify its environment in a process known as niche construction.  
This feedback loop exemplifies compositional evolution, where the cell and environment co-adapt.
Cells actively reshape their surroundings by secreting extracellular matrix components, exerting mechanical forces, adhering to surfaces, and releasing enzymes such as proteases and hydrolases. 
They redistribute nutrients and signaling molecules, creating sensorimotor and chemical feedback loops. 
These interactions allow cells to construct favorable niches, enabling adaptation to changing conditions and achieving desired states through their interface.

At this point in the narrative, our composite models of cells and environments may resemble ABMs, which emphasize a cell’s behavioral responses to its surroundings, simplifying or ignoring the underlying molecular processes. 
For instance, an ABM might establish rules for a cellular update function, such as moving towards or consuming nutrients based on the cell’s internal state of hunger and environmental cues. 
Decision-based models enhance this by incorporating a logic that prioritizes actions based on their utility—a configuration that maps a cell’s input states to some value or reward that needs to be maximized, such as growth, robustness, or replication. 
However, being more phenomenological, these models fall short in detailing the molecular mechanisms that drive a cell's behaviors, functions, or decision-making processes, and they do not explain the fundamental reasons why a cell \textit{needs} nutrients in the first place. 
To delve deeper into these aspects, we need to shift to a molecular perspective.

\subsection*{Molecular interface}
When we peel away the membrane, we see the cell is itself a composite system of many interacting molecules, which can become associated together in assemblies such as complexes, condensates, and organelles. 
Molecules interact with each other through various mechanisms, including chemical reactions, binding, transport, electrostatic interactions, mechanical forces, and so forth. 
To understand how these interactions compose into cellular functions, we first need to characterize the molecular interface.

A molecule can engage with other molecules in many ways, but contemporary models often focus narrowly on a subset of this interface. 
For instance, molecular models may focus on the kinetics of a reaction catalyzed by a specific protein, while ignoring its structural properties or spatial location. 
Reducing the molecular interface simplifies the resulting model and enhances computational efficiency, but also limits its scope and ability to compose in the same way real molecules do. 
A compositional framework expands these limitations by integrating multiple modeling paradigms into a unified molecular interface.
There has already been substantial progress in integrative modeling for the structure of molecules \cite{sali2021integrative}—we also need to integrate functional models.
Composition could bring together structural models with molecular dynamics, kinetic reaction networks, particle-based stochastic models, coarse-grained models, continuum models, etc. 
Each of these frameworks provides unique insights and can be composed to reflect the complex interplay of molecular interactions within the cellular environment.

\begin{figure}[H]
    \centering
    \includegraphics[width=1.0\textwidth]{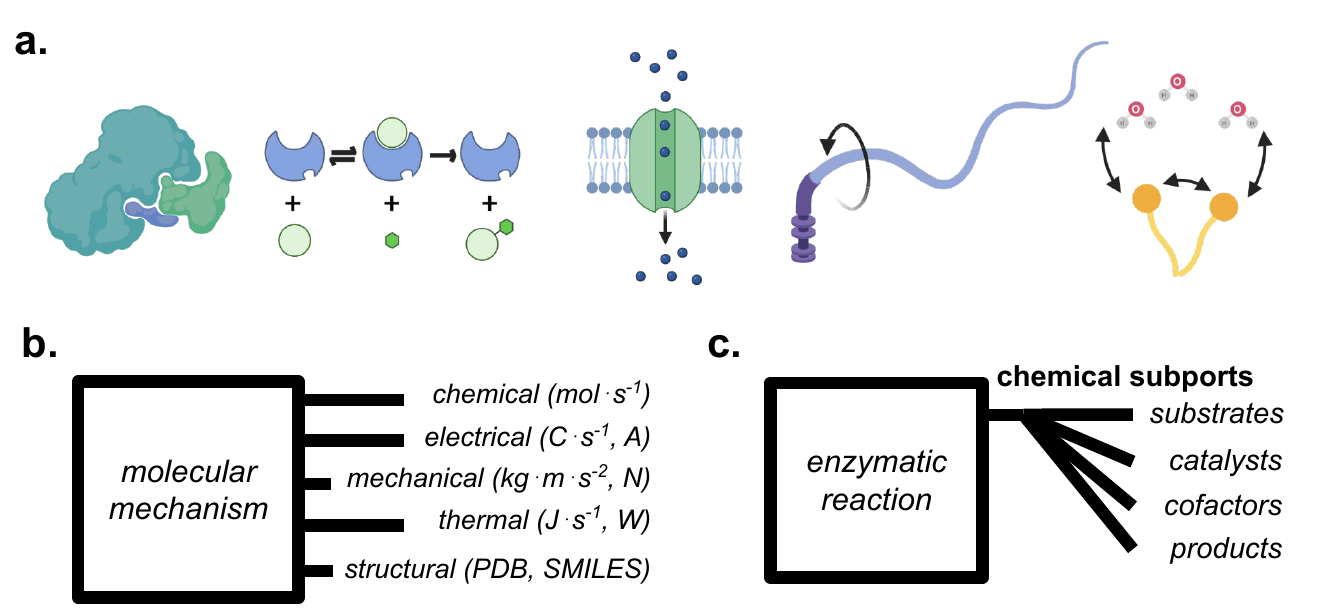}
    \caption{
    \textbf{Molecular mechanisms as processes that update molecular states.}
    \textbf{a.} These can take a range of forms: a structural model of how molecules fit together, enzymatic reaction with an enzyme attaching two small molecules, a transmembrane transport process, motility-generation with a flagella's rotation-driven thrust, and lipid hydrophobic and hydrophilic interactions.
    \textbf{b.} A molecular interface diagram showing basic molecular properties with their own types and constraints: chemical, electrical, mechanical and thermal ports.
    \textbf{c.} A more specialized molecular mechanism could subdivide any of the ports, such as an enzymatic reaction differentiating between the substrates and cofactors that are taken as input into the reaction, a catalyst, and the resulting products and changes to the substrates as outputs.
    }
    \label{fig:molecular_interface}
\end{figure}

Ports can be further subdivided according to the molecular mechanisms. 
For instance, enzymatic reactions (Fig. \ref{fig:molecular_interface}c) are specialized processes that divide chemicals into well-defined subtypes: 1) substrates: molecules that enzymes act upon, 2) catalysts: enzymes or other molecules that speed up reactions, 3) cofactors: non-protein molecules that assist enzymes, 4) products: molecules produced from enzymatic reactions.

A comprehensive molecular interface defines interactions as processes with ports (Fig. \ref{fig:molecular_interface}), encompassing chemical reactions, physical forces like van der Waals and electrostatic interactions, hydrophobic effects critical for protein folding, and steric influences on molecular shape and reactivity. 
Additional ports account for conditions such as temperature, pH, or mechanical forces, enabling dynamic modulation. 
Formalizing these interfaces integrates structural, kinetic, and spatial dimensions, supporting the composition of molecular systems into higher-order functions and forming the basis for the intricate networks that sustain cellular life.

\subsection*{Biomolecular assemblies}
\begin{figure}[H]
    \centering
    \includegraphics[width=1.0\textwidth]{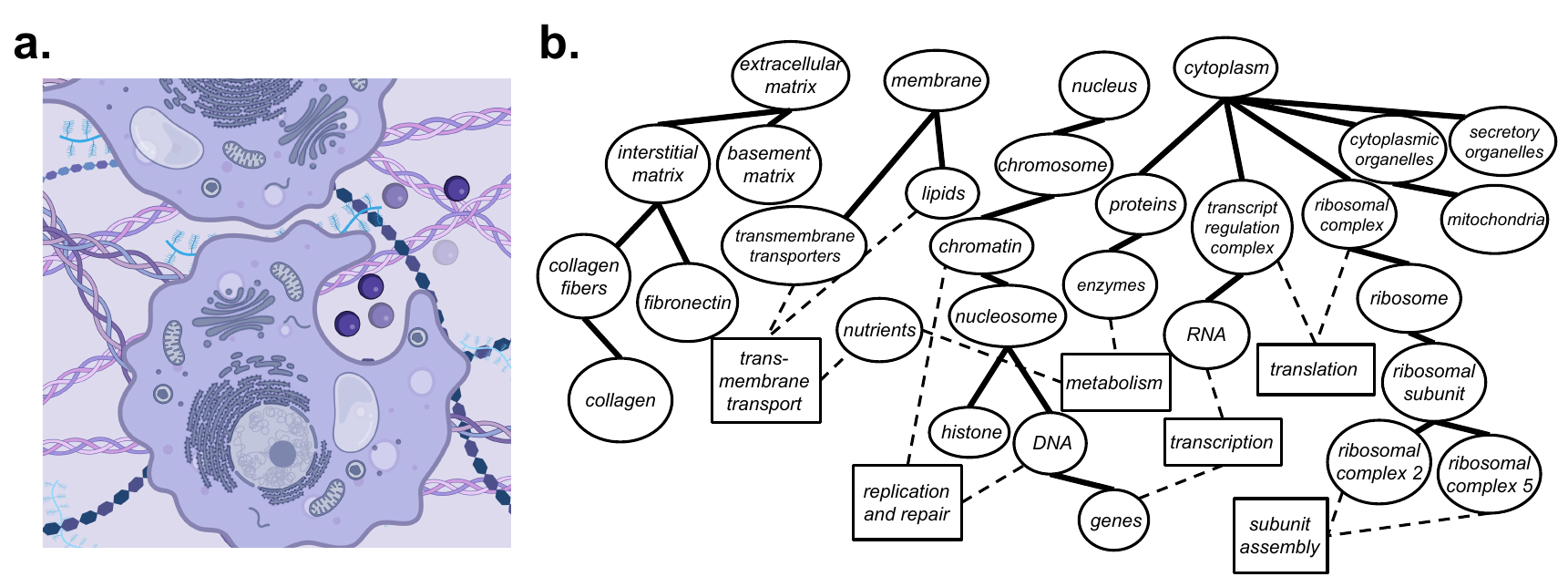}
    \caption{
    \textbf{Molecular compositions as nested hierarchical composites.}
    \textbf{a.} Different molecular structures both within and external to a cell, including a membrane, organelles, nucleus, and extracellular matrix. 
    Complexes such as the ribosome complex can be subdivided into smaller complexes, all the way down to individual proteins.
    \textbf{b.} Composite processes diagram showing nested structure of a molecular assemblies, with circles representing nested states and rectangles representing processes. 
    Each process, such as transmembrane transport, transcription regulation, and metablism connects with specific nested states like the membrane, cytoplasm, and nucleus. 
    These interactions facilitate various cellular functions.
    However, without organizational principles for the cellular interface, this process bigraph can easily become cluttered and hard to interpret.
    }
    \label{fig:molecular_compositions}
\end{figure}

We can begin assembling molecular processes through shared, complementary states, creating larger and more intricate nested structures made of many molecules (Fig. \ref{fig:molecular_compositions}). 
There are many ways to approach this task—it can include the progressive composition of metabolic pathways \cite{braakman2012compositional}, lipids \cite{segre2001lipid}, or aggregates of different complementary molecules \cite{hunding2006compositional}; it could take integrating the spatial distribution of proteins with their interactions \cite{qin2021multi}, or taking individual molecular structures and packing them together into a volume \cite{johnson2015cellpack}.
Each molecular process is defined by an interface with specific ports—such as active sites on enzymes, binding sites on receptors, and docking domains on structural proteins—which facilitate the coalescence of molecules into functional complexes based on the complementarity of their interfaces. 
These interactions are orchestrated through a set of molecular affinities and repulsions, aligning and connecting molecules to based on functional activity. 
As these complexes integrate into larger compositions, they can form condensates through mechanisms such as liquid phase separation \cite{banani2017biomolecular}. 
Spatial segregation by membranes encapsulates these assemblies into organelles, creating subcompartments with distinct internal molecular states and boundary conditions. 

As we approach the scale of a cell, the challenge of assembling to molecular mechanisms in a way that established a cellular interface from molecular components becomes evident, as the boundary between internal and external states remains ambiguous, the required resolution is undefined, and the organization of these states is unconstrained.
Take the example illustrated in Fig. \ref{fig:molecular_compositions}b—identifying a cellular interface in this hierchical molecular composition would take drawing a process rectangle around a subset of the states that can specify a minimal cellular interface, delineating the molecules and processes inside of the cell from those outside of a cell. 
Molecular mechanisms—whether transmembrane transport, gene expression, protein synthesis, signal transduction, metabolism, protein folding, or DNA replication—lack a clear distinction between whether it is inside or outside of a cell. 
Processes such as transport reactions, membrane trafficking, and the production of extracellular matrix can link across what we conventionally consider the cell boundary. 
Despite our ability to assemble molecular components into large-scale composite structures, systems biology still struggles to define what makes one particular molecular composition a living cell and the other one dead matter. 
Without a clear framework to distinguish living molecular compositions from non-living molecular compositions, we cannot objectively determine which individual we are observing or how it adapts within its environment to meet its needs.

\subsection*{Self-organization, coarse-graining, and the emergence of the cellular interface}

\begin{figure}[H]
    \centering
    \includegraphics[width=1.0\textwidth]{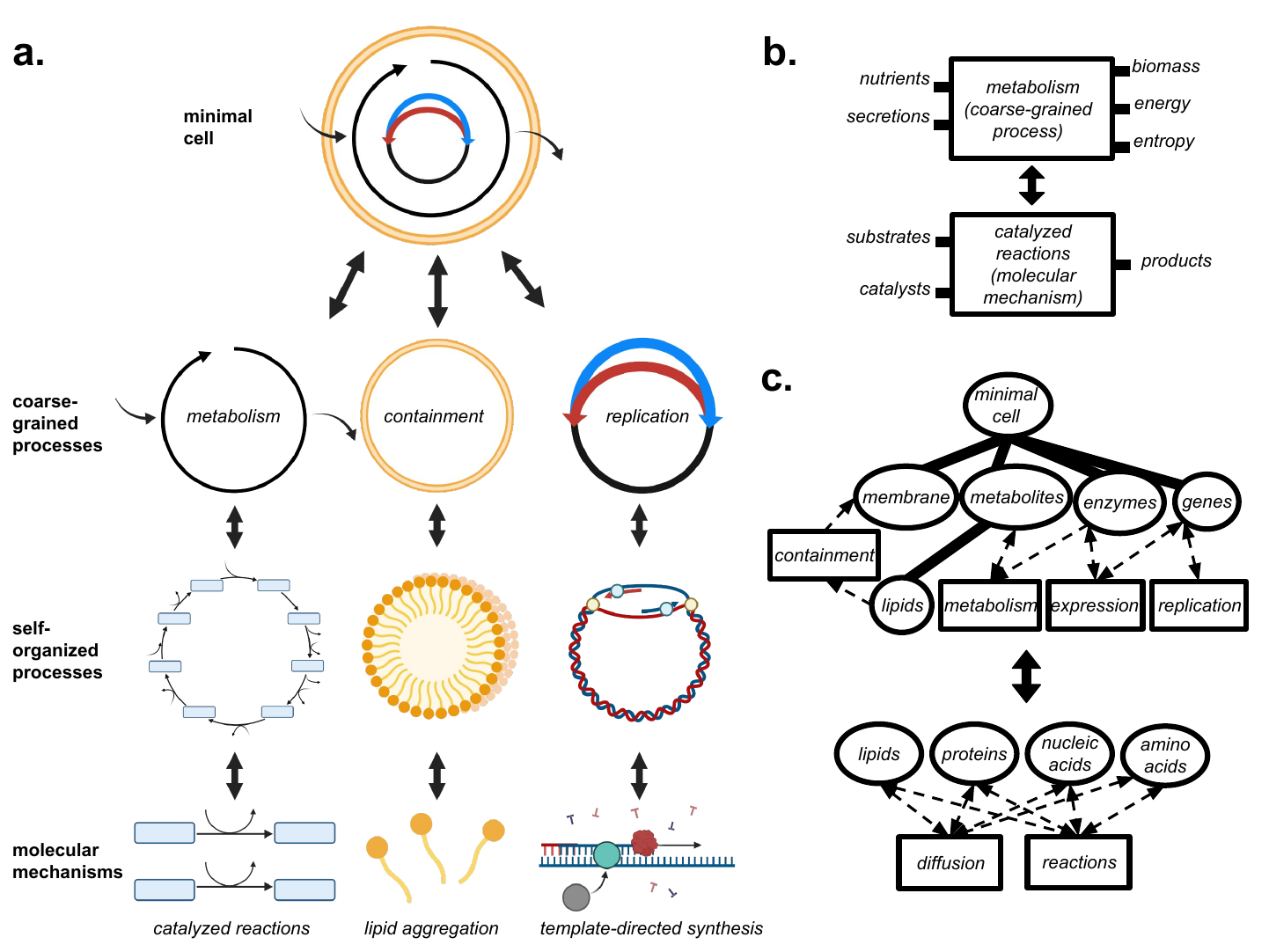}
    \caption{
    \textbf{Self-organized processes, coarse-graining, and the minimal cell composition.}
    \textbf{a.} Several self-organized processes: autocatalysis, membranes, and template replication compose into a self-sustaining autopoietic organization.
    \textbf{b.} Diagram showing the swapping of spontaneous catalyzed reactions, a molecular process, with a coarse-grained metabolism process that captures several catalyzed reactions including loops. The composite would need to identify the emergence of a closed metabolic loop within the molecular simulation, and replace it with the coarse-grained processes. 
    \textbf{c.} A minimal cell composition diagram with minimal required metabolism, containment, and replication processes, supporting each other through catalysis of required parts, containment to counter diffusion, and replication to support the generation of new minimal cells. 
    }
    \label{fig:self_organized_process}
\end{figure}

A key challenge in multiscale modeling is determining when to coarse-grain complex molecular processes and when to revert to more detailed representations (Fig. \ref{fig:self_organized_process}). 
Coarse-grained processes allow for more intuitive modeling at a scale that matches experimental observations, can integrate macroscopic data types, and is computationally less-expensive.
This decision of when to coarse-grain is crucial for capturing the dynamics of self-organized biomolecular systems, which emerge from independent molecular processes and form higher-level structures with new properties.

Self-organized processes arise when molecular compositions harness external energy gradients to maintain order far from equilibrium \cite{morowitz1979energy}. 
When such processes form, we may want to identify them as emergent self-organized processes that have their own properties, and abstract away the underlying molecular processes and extra degrees of freedom from which they emerge. 
A self-organized process would have a thermal port, to release heat into the environment, and a work energy port, which taps into external gradients to drive sustained nonequilibrium organization. 
As the external gradient depletes, the non-equilbrium molecular system decays back to equilibrium and the self-organized process ceases to exist, replaced again by independent molecular processes. 

Autopoiesis is a unique form of composition where two or more self-organized processes keep each other active by maintaining each others required gradients, enabling continuous self-construction. 
Processes in an autopoietic composition form a network of self-organized molecular processes that continually re-synthesize their required materials and gradients and maintains a boundary that separates them from their environment \cite{maturana1991autopoiesis}.
Metabolic networks (Fig. \ref{fig:self_organized_process}a, top) counter molecular degradation through autocatalysis, while self-assembled lipid membranes containers (Fig. \ref{fig:self_organized_process}A, middle) counteract the entropic effects of diffusion—together they can counter degradation and diffusion by maintaining an active compartmentalized metabolism.
Closed membranes also allow for proton-motive force and other chemical gradients to be maintained by the system.
Template replication (Fig. \ref{fig:self_organized_process}a, bottom) is a process by which a template of DNA or RNA is copied.
Their mutual support enables these cellular processes to self-organize as distinguishable and adaptive entities separate from their environments.
This again raises the question of a minimal cellular interface—what are the minimal sets of self-organized molecular processes that can together recreate a cellular interface?

Coarse-graining must balance preserving the integrity of cellular models with the need to revert to molecular detail when systems fail. 
While stable cells can be abstracted as coarse-grained entities, this risks obscuring the precarious nature of life and the entropic forces driving disintegration \cite{beer2023theoretical}. 
Cellular models must capture the dynamic transitions between organization and decay, allowing them to reflect the critical processes of both self-organization and disintegration.

\subsection*{Growth, division, development, evolution}
\label{sec:grow_divide}

\begin{figure}[H]
    \centering
    \includegraphics[width=1.0\textwidth]{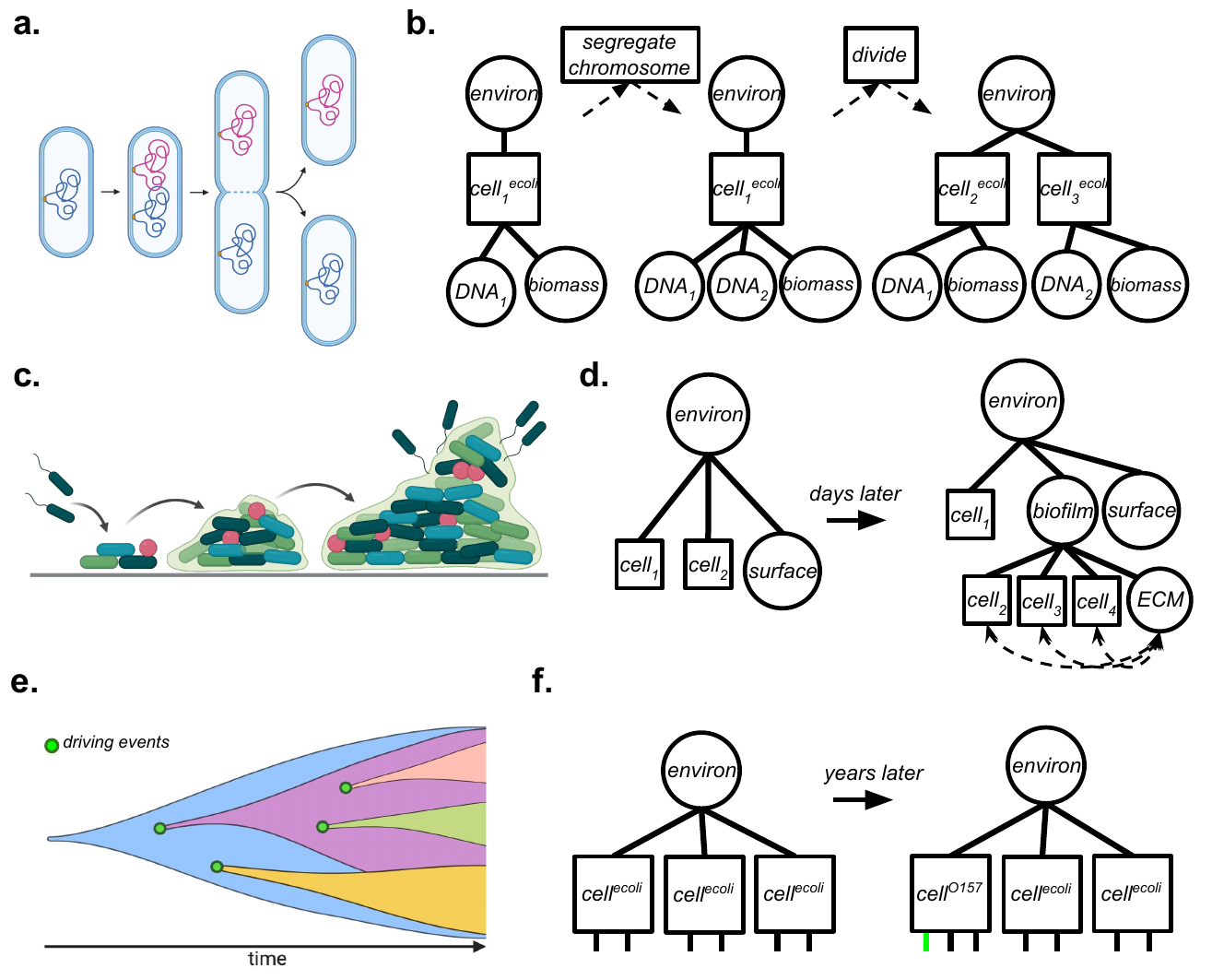}
    \caption{
    \textbf{Growth and division leading to development and evolution.}
    \textbf{a.} Prokaryotic cell division. The cell initiates replication of DNA as cell elongates, the cross wall forms, and daughter cells separate. 
    \textbf{b.} Diagram of cell division as hierarchical updates, starting one cell before that has some DNA and biomass, resulting in two cells afterwards with a new copy of the DNA and each with its own biomass.
    \textbf{c.} Biofilm development, initiated by free-living bacteria that attach to a surface, secrete ECM, and grow with functional specialization.
    \textbf{d.} The development of biofilm includes a hierarchical change, in which free-living bacteria come under a unified biofilm composition in which multiple cells are bound together by an extracellular matrix (ECM).
    This transition in the process bigraph would require a biofilm process that recognizes when biofilm activities are present and moves the cells to a new biofilm store.
    \textbf{e.} Evolution over time, with key driving events shown in green, leading to new functionality that expands in the population over time.
    \textbf{f.} The driving event may come in the form of a new interface port added to a cell, shown in green. 
    This can give the cell new ability to sense or influence the environment, leading to evolutionary expansion.
    }
    \label{fig:divide_evolve}
\end{figure}

As cells absorb nutrients from their environment and convert them into biomass, they grow in mass, volume, and surface area. 
This alters their surface-to-volume ratio, a key variable affecting how efficiently they interact with their surroundings.
Rod-shaped bacteria, for example, maintain a stable ratio by elongating while keeping a fixed width (Fig. \ref{fig:divide_evolve}a,b). 
Growth also drives compositional changes within the cell as new proteins, lipids, and metabolites are synthesized, diluting existing molecules and reshaping cellular composition and functionality \cite{taymaz2013changes,schmidt2016quantitative,mori2021coarse}.

Following growth, division marks a topological transformation where one cell bifurcates into two. 
During division, the membrane and cytoplasm reorganize, partitioning molecules into two unique daughter cells, each with its own interface for interacting with the environment. 
These interfaces often remain coupled due to spatial proximity, influencing each other and forming the foundation for heterogeneous cellular assemblies that interact with diverse environmental pressures.

There are different ways cellular interfaces can interconnect with each other through processes of adhesion, predation, interaction, communication; these can range from simple interactions between two bacterial cells to complex ecosystems with many cells, including organisms made of many specialized cells. 
Symbioses include mutualistic, commensalistic, or parasitic interactions, forming stable compositions of cellular populations. 
As an example in bacteria, the development of multicellular structures seen in biofilms (Fig. \ref{fig:divide_evolve}c,d), where communal life is distinct from that of free-living bacteria \cite{stewart2008physiological,flemming2016biofilms,kostakioti2013bacterial}, demonstrates how a multicellular composite exhibits emergent properties such as cooperation, resource capture, and increased survival under stress.
The cells produce a physical scaffold of extracellular matrix, holding them together and attaching them to environmental surfaces, and driving physiological changes to their  populations.

As cellular populations grow and divide, their heterogeneity increases due to stochastic processes both within and outside the cells (Fig. \ref{fig:divide_evolve}e,f). 
Efficient cellular interfaces that optimize resource use, enable communication, and withstand environmental pressures are naturally selected, while less viable cells are eliminated. 
Multicellular compositions dynamically restructure to optimize energy flow and resource allocation, integrating new processes to enhance functionality and resilience.
Orchestration within these systems enables the rearrangement and recombination of processes, allowing the exploration of novel configurations and pathways. This iterative refinement, driven by environmental pressures, expands the functional complexity of the system, creating more efficient and adaptive structures.
Through these adaptive interactions and emergent properties, multicellular compositions evolve into increasingly sophisticated systems, continually pushing the boundaries of evolutionary innovation.

%%%%%%%%%%%%%%%%
% Collaborative biosciences  %
%%%%%%%%%%%%%%%%

\section*{Collaborative biosciences}
\label{sec:collaborative_biosciences}

The composition framework and biological templates introduced in this article establish a foundation for modeling multiscale cellular systems with unprecedented detail across molecular, cellular, and multicellular scales. 
Central to this effort is the larger composite system of scientific, social, and technological tools developed for biological model building and verification. 
This system, orchestrated by human researchers, integrates experimental tools, computational technologies, and biological systems to iteratively refine and expand our understanding of cellular systems.

Scientific collectives—teams of researchers, datasets, and models—function as compositions interconnected by shared resources like scientific papers, databases, and software libraries. 
These collectives operate as integrated systems, driven by human ingenuity to connect diverse knowledge domains. 
Just as cells rely on coordinated molecular processes, scientific collectives depend on collaboration to construct, validate, and refine realistic, accessible models.

Enhancing researcher productivity and integrating computational models with real-world systems requires a human-centered approach. 
Intuitive tools like Jupyter notebooks, graphical user interfaces (GUIs), and AI-powered assistants streamline data navigation, model updates, and prediction validation. 
Digital twins establish feedback loops that continuously improve models through streaming updates \cite{laubenbacher2022building}. 
AI complements these tools by automating tasks, navigating datasets, and highlighting data gaps through uncertainty quantification \cite{johnson2023building}. 
Gamification strategies further engage researchers, transforming modeling into interactive, rewarding experiences \cite{das2019scientific}. 
By incorporating human-in-the-loop approaches, these systems empower researchers to focus on creative problem-solving while maintaining alignment with scientific workflows.

On an organizational scale, effective collaboration strategies must enable seamless integration of data and models, allowing researchers to ask advanced, integrative questions. 
A shared type system linking modeling and experimental efforts is essential, defining consistent formats and standards to ensure experimental outputs directly inform simulations. 
Tools like GitHub facilitate version control and iterative model evolution, while shared platforms improve accessibility and interdisciplinary participation. 
An open composition standard, as detailed in Section \ref{sec:standard_composition_protocol}, ensures interoperability between tools and systems, enabling custom interfaces and seamless integration.

A compositional framework for systems biology can redefine how we address the complexities of cellular systems, emphasizing integration across data, models, human expertise, and technological tools. 
By establishing shared standards and fostering modularity, this approach bridges modeling and experimentation, enabling iterative, bidirectional refinement. 
Scientific progress becomes a collective, multiscale effort where every model, dataset, and insight serves as a modular component in a larger scientific architecture. 
Compositional systems biology thrives to transform isolated research efforts into interconnected systems of discovery, creating ecosystems capable of tackling intricate biological questions and driving transformative breakthroughs.

%%%%%%%%%%%%
% Acknowledgements  %
%%%%%%%%%%%%

\section*{Acknowledgements}
Thank you to my many colleagues for the discussions that seeded and developed the ideas behind this article, especially to Ryan Spangler for co-developing the process bigraph framework and software. 
E.A. is funded by NSF award OCE-2019589 to the Center for Chemical Currencies of a Microbial Planet, by NIH award P41GM109824 to the Center for Reproducible Biomedical Modeling, and John Templeton Foundation award \#62825.
This is C-CoMP publication \#48.
We thank BioRender.com for providing the tools to create the scientific illustrations used in this paper.

\section*{Author Contributions}
E.A. conceived the framework and wrote the manuscript. 

\section*{Competing Interests}
The authors declare no competing interests.

\section*{Data Availability}
No datasets were generated or analyzed during this study.

\section*{Code Availability}
No code were created during this study.

\printbibliography

\end{document}